\documentclass[a4wide,twocolumn,11pt]{article}

\usepackage[english]{babel}
\usepackage{amsmath,amssymb,graphicx}
\usepackage[tikz]{mdframed}
\usepackage{a4wide}
\usepackage{hyperref}

\newcommand{\nobracket}{}
\newcommand{\tmem}[1]{{\em #1\/}}
\newcommand{\tmmathbf}[1]{\ensuremath{\boldsymbol{#1}}}
\newcommand{\tmop}[1]{\ensuremath{\operatorname{#1}}}
\newcommand{\tmstrong}[1]{\textbf{#1}}
\mdfsetup{linecolor=black,linewidth=0.5pt,skipabove=0.5em,skipbelow=0.5em,hidealllines=true,innerleftmargin=0pt,innerrightmargin=0pt,innertopmargin=0pt,innerbottommargin=0pt}
\newmdenv[hidealllines=false,innertopmargin=1ex,innerbottommargin=1ex,innerleftmargin=1ex,innerrightmargin=1ex]{tmornamented}

\makeatletter
\def\blfootnote{\xdef\@thefnmark{}\@footnotetext}
\makeatother

\begin{document}
\twocolumn[
\title{The comprehensive theory of light}
\author{\textsc{Urjit A. Yajnik}
\\
\textsl{Indian Institute of Technology Bombay}
}
\date{}
\maketitle
]
\section{A persistent dichotomy}\blfootnote{Published in \textsl{Physics News} Bulletin of Indian Physics Association, vol. \textbf{49}, No. 1, 2019}\label{sec1}

Modern attempts to understand light go back to Newton who considered light to
be particles, the so called corpuscular theory, and the other school of
Huygens, Young and others. Huygens and Young viewpoint emphasised the wave
property. This ``difference of opinions'' persisted for close to two centuries
till Maxwell theory solidly established light as a wave phenomenon associated
with Electromagnetism. The invention of \ transmitters and receivers of
electromagnetic waves to which J C Bose made ingenious contributions
established \ electromagnetic waves on a firm footing.

A serious schism so to speak was introduced into theory of light with the
understanding of ``Light gas'', the so called Black Body radiation. Planck
could develop a theory for the spectral distribution of cavity radiation only
by associating quantum properties to the processes of absorption and emission
of light by the oscillators in the walls of the cavity. It was young Einstein
a few years later who could see clearly what was going on. The quantum
property was not a peculiarity of the oscillators. It was a fundamental
property of light itself. He went on to formulate this atomistic hypothesis
about light quantum as :
\begin{quotation}
  ``According to the assumption considered here, in the propagation of a light
  ray emitted from a point source, the energy is not distributed continuously
  over ever-increasing volumes of space, but consists of finite number of
  energy quanta localised at points of space that move without dividing, and
  can be absorbed and generated only as complete units''.
\end{quotation}
Einstein's hypothesis helped him to deduce the equation of the photoelectric
effect. But his viewpoint took close to two decades to be accepted. And this
happened only after S. N. Bose provided the key derivation of Planck's formula
based on this hypothesis, and using Boltzmann's method of
distributions[\ref{ref:uayphotonsonetwo}]. But by this time, even the
conception regarding the nature of matter was encountering a major dichotomy :
electrons could appear to be particles or waves, depending on the
circumstances. Likewise light now had this dual behaviour, that in waves
versus that in Photoelectric effect, Compton effect, X-rays etc.

However for all practical purposes, electromagnetism continued to have the
completely classical description as Maxwell's waves and the latter enjoyed
complete success as the description of electromagnetism in all walks of
engineering. One may then wonder if this a classical limit may have to be
corrected when dealing with quantum phenomena. To most people's surprise, the
classical description was in fact a subset of the full quantum description,
and the classical states of light could be shown to be subsumed within the
fully quantum description without having to take an $\hbar \rightarrow 0$
limit. While the exact correspondence is technical due to the use of complex
number notation, this was in effect the resolution provided by Sudarshan's
Diagonal representation as the most general formalism for dealing with light.
\

\section{Approaches to comprehensive description of light}

The Huyghens and Young school viewpoint comprehensively covered all the
macroscopic phenomena. It was corroborated by a variety of experiments and
indeed, ray optics could be derived from it. The need to develop a more
detailed theory of light arose with Astrophysical observations of Hanbury
Brown and Twiss in the late 1950's. Two new issues to be faced were : (i) the
intensities (photon counts) in two distant detectors were correlated even when
the phase relationship had been lost. (ii) The role of statistics, which at
first appeared to be entirely deducible from Maxwell theory and stochastic
effects, could not however be adequately explained. In order to understand
these developments it is useful to understand the concepts in use for the
classical description of statistical properties of radiation as developed in
the work of Emil Wolf and others. In this one defines ``coherence functions''
which are correlation functions of components of the electric field at
different points. Thus the ``second order coherence function'' is defined as
\begin{eqnarray*}
  \Gamma (\tmmathbf{x}_i t_i, \tmmathbf{x}_j t_j) & = & \langle V^{\ast}
  (\tmmathbf{x}_i, t_i) V (\tmmathbf{x}_j, t_j) \rangle\\
  & \equiv & \Gamma (\tmmathbf{x}_i, \tmmathbf{x}_j, \tau) \label{seccohfn}
\end{eqnarray*}
where $\tau = t_i - t_j$, and the $V$ is the positive frequency part to be
extracted from the temporal Fourier transform of the electric field and
referred to as ``analytic signal''. Higher order coherence functions could be
similarly defined, involving field strength at several space time points.
Since intensity is determined by square of the local value of the electric
field, that information is contained in such functions.

\

A theorem that connects the coherence function to observables and makes this
formalism tractable can be derived if one introduces the more restricted
quantity ``reduced coherence function''
\[ \gamma (\tmmathbf{x}_i t_i, \tmmathbf{x}_j t_j) = \frac{\Gamma
   (\tmmathbf{x}_i t_i, \tmmathbf{x}_j t_j)}{\sqrt{\Gamma (\tmmathbf{x}_i t_i,
   \tmmathbf{x}_i t_i) \Gamma (\tmmathbf{x}_j t_j, \tmmathbf{x}_j t_j)}} \]
It can be shown [\ref{ref:ecgreview}] that this quantity has direct
interpretation as the ``visibility index'' or the contrast between the\quad
intensity maxima and minima. The coherence functions can now be shown to have
two important properties, viz., if $\Gamma_{(k)} (\tmmathbf{x}_i t_i,
\tmmathbf{x}_j t_j)$ is a set of valid coherence functions then so is
\[ \Gamma (\tmmathbf{x}_i t_i, \tmmathbf{x}_j t_j) = \sum_k \lambda_k
   \Gamma_{(k)} (\tmmathbf{x}_i t_i, \tmmathbf{x}_j t_j) \]
provided $\lambda_k$ are nonnegative numbers. This property is called
convexity. Further the set of all such functions, said to constitute a convex
cone have as their generators those coherence functions whose corresponding
reduced coherence functions are unimodular, i.e., $| \gamma | = 1$.

This introduction prepares us to understand the stage at which Sudarshan
entered the field. It was realised by early 1960's that purely classical and
stochastic or thermal effects although sufficient to understand the outcome of
Hanbury Brown and Twiss type experiments qualitatively could not account for
magnitude of the effect. A formalism based on quantum mechanics was developed
by Roy J. Glauber who introduced coherent states, which can be understood as
the eigenstates of the quantum analogues of the analytic signals being used in
the classical formalism.

\begin{center}
\begin{figure}[ht]
\includegraphics[width=0.9\columnwidth]{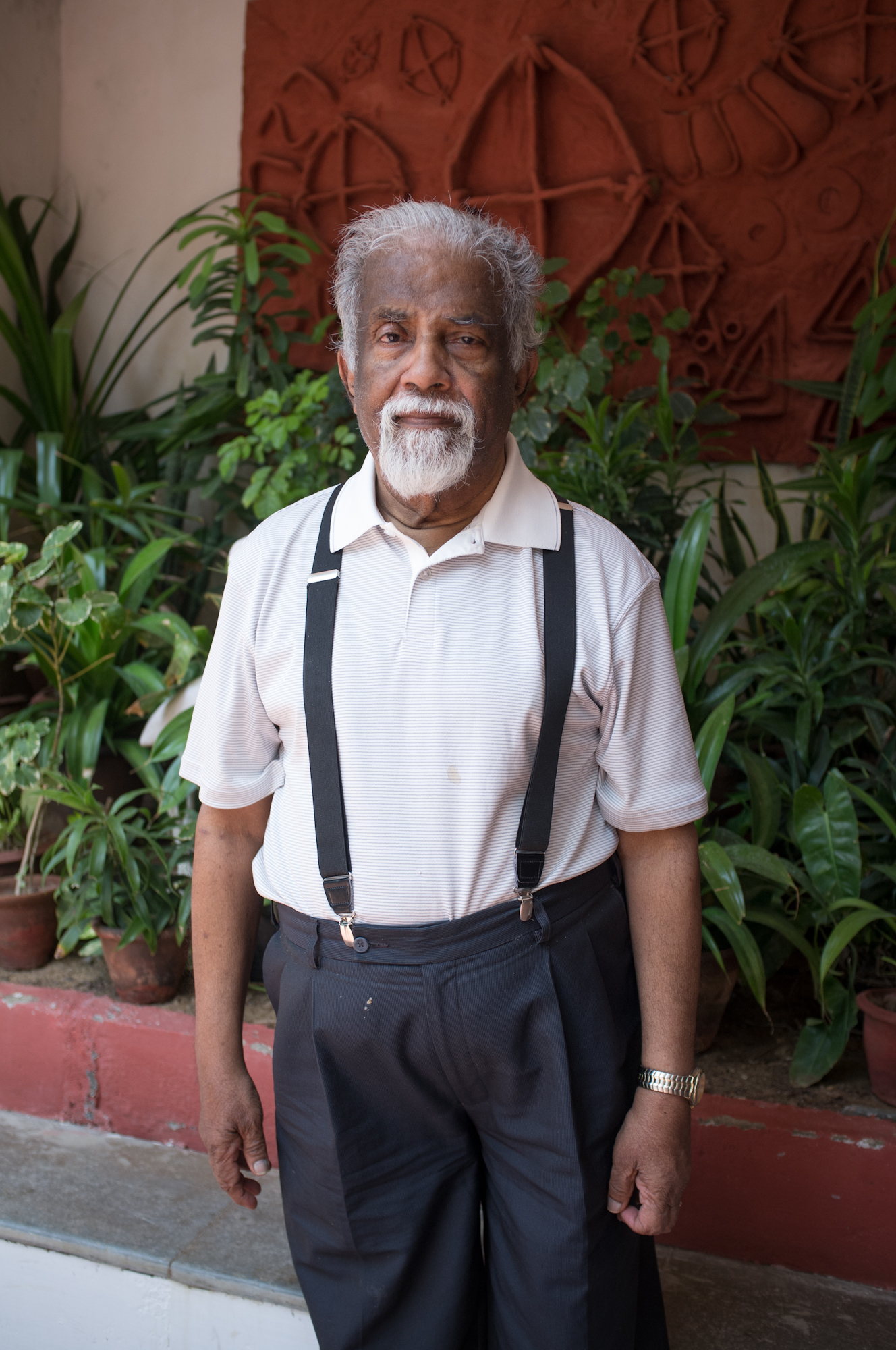}
\newline
\centering{E. C. G. Sudarshan in August 2014 \newline{\small Photo courtsey Avani Tanya}
}
\end{figure}
\end{center}

\section{Coherent states}

Let us consider a harmonic oscillator, with Hamiltonian given in terms of the
canonical variables as
\[ H = \frac{p^2}{2 m} + \frac{1}{2} m \omega^2 q^2 \]
When we quantise this system by imposing
\[ [q, p] = i \hbar \]
we find that the Hamiltonian has the spectrum of eigenvalues
\[ E_n = \left( n + \frac{1}{2} \right) \hbar \omega . \]
An elegant way to obtain this result is to introduce the creation and
destruction operators. For convenience of focusing on the essentials we set
$m$ and $\omega$ to unity and also set $\hbar = 1$. Then one introduces
\[ a = \frac{1}{\sqrt{2}} (q + i p), \qquad a^{\dag} = \frac{1}{\sqrt{2}} (q
   - i p) \]
which serve as the analogues of the classical analytic signals. We see that
they satisfy
\[ [a, a^{\dag}] = 1 \]
and we recover the Hamiltonian as
\[ H = a^{\dag} a + \frac{1}{2} \]
with eigenstates $| n \rangle \nobracket$ obtainable as
\[ | \nobracket n \rangle = \frac{1}{\sqrt{n!}} (a^{\dag})^n | \nobracket 0
   \rangle ; \qquad a | \nobracket 0 \rangle = 0 \]
As all students of quantum mechanics know, the states labeled by $n$ are
completely counter intuitive from the point of view of classical mechanics. On
other hand, if we seek states that \ accord more closely to classical
intuition, they are the ``minimum uncertainty wave packets''. These are the
states which saturate the uncertainty principle statement
\[ \Delta q \Delta p \geqslant \frac{1}{2} \]
i.e. when the uncertainties are evaluated in coherent states, the above
statement becomes an equality. E. Merzbacher's textbook [\ref{ref:merz}]
contains a detailed discussion.

While harmonic oscillator is not essential to understand coherent states, it
provides a bridge to motivate the introduction of the creation and destruction
operators. Also, the free electromagnetic field has a hamiltonian essentially
of the same type. We next discuss the essential properties of these states, a
discussion largely based on [\ref{ref:mukunda}][\ref{ref:mehta}].
\begin{enumerate}
  \item Coherent states are defined as eigenstates of the destruction operator
  \[ a|z \rangle = z|z \rangle \]
  The eigenvalues $z$ are complex, which is not surprising since $a$ is not
  hermitian. Below we shall see the connection of these complex numbers to the
  ordinary coordinates $q$ and $p$.
  
  \item Coherent states are not orthogonal.
  \[ \langle z' | z \rangle = \exp \left\{ - i \tmop{Im} (z' z^{\ast}) -
     \frac{1}{2} | z' - z |^2 \right\} \]
  \item There is a ``resolution of the identity'',
  \[ \int \frac{d^2 z}{\pi} | z \rangle \langle z | = 1 \]
  but this is not a ``completeness relation'' due to lack of orthogonality. In
  fact they are overcomplete as a basis set.
  
  \item Displacement operator : Consider $z$ written as $(q + i p) / \sqrt{2}$
  where $q$ and $p$ are any real numbers. But we shall next see that these
  also have the interpretation of belonging to the set of real eigenvalues of
  the operators $q$ and $p$. Let
  \begin{eqnarray*}
    D (z) & \equiv & D (q, p)\\
    & = & \exp (z a^{\dag} - z^{\ast} a)\\
    & = & \exp \{ i (p \hat{q} - q \hat{p}) \}
  \end{eqnarray*}
  where we have temporarily put hats on the $p$ and $q$ to distinguish
  operators from eigenvalues. Further,
  \[ D (z) D^{\dag} (z) = 1 \]
  Then we find that
  \begin{eqnarray*}
    D (z)^{- 1} \hat{q} D (z) & = & \hat{q} + q\\
    D (z)^{- 1} \hat{a} D (z) & = & \hat{a} + z
  \end{eqnarray*}
  and similar relations for $\hat{p}$ and $\hat{a}^{\dag}$. This justifies the
  name displacement operator, and the interpretation of $\tmop{Re} z$ as a
  possible eigenvalue of $\hat{q}$ and that of $\tmop{Im} z$ as that of
  $\hat{p}$.
  
  From this, we can also interpret the states $| z \rangle \nobracket$ as
  simply displaced versions of the vacuum
  \[ | z \rangle \nobracket = D (z) | 0 \rangle \nobracket \]
\end{enumerate}
This ends the list of four essential constructs and facts to be introduced. We
may now consider using the set $\{ | z \rangle \}$ as a basis. Due to the
resolution of identity, we can express any state $| \psi \rangle \nobracket$
as
\[ | \psi \rangle \nobracket = \int \frac{d^2 z}{\pi} | z \rangle \langle z |
   \psi \rangle \]
However the wave function $\psi (z) \equiv \langle \nobracket z | \psi
\rangle$ is not unique due to lack of orthogonality. We have for example,
\[ \int z | z \rangle \nobracket d^2 z = 0 \]
so $z$ can be added to any wavefunction without affecting $| \psi \nobracket
\rangle$.

In quantum mechanics, the density matrix formalism is also convenient. Given a
state vector \ $| \psi \nobracket \rangle = \sum c_n | n \nobracket \rangle$
the same amount of information is encoded in the operator
\[ \rho = | \psi \rangle \langle \psi | \nobracket = \sum_{n, n'}
   c^{\ast}_{n'} c_n | n \rangle \langle n' | \nobracket \equiv \sum_{n, n'}
   \rho_{n n'} | n \rangle \langle n' | \nobracket \]
In turn we may consider $| n \rangle \langle n' | \nobracket$ as the basis for
expressing such operators. Here we used the orthonormal basis $\{ | \nobracket
n \rangle \}$ and the representation $\rho_{n n'}$ is unique. However consider
proposing a similar representation in the $\{ | z \rangle \}$ basis,
\[ \rho = \int d^2 z F (z, z') | z \rangle \langle z' | \nobracket \]
Such a representation $F (z, z')$ is not unique, in \ terms of each of its
arguments.

\section{A forfeited lunch}

Sudarshan has recalled in his memoirs [\ref{ref:ecgftof}] that he had an in
depth exposure to optics from an excellent teacher Mr. Thangaraj at Madras
Christian College. This must have certainly helped him to communicate and
discuss the subject with pioneers and experts like Emil Wolf and Leonard
Mandel when he went to Rochester. He recalls that in 1963 Wolf returned from
Les Houches workshop in Europe where Glauber had given a set of \ lectures
introducing the fully quantum treatment of Hanbury Brown and Twiss
experiments, and had also specifically said that the classical treatment of
the subject needs to be abandoned. Wolf was rather dejected as that was his
life's work, and he despaired at having to learn quantum mechanics. Sudarshan
set out to reassure him that the quantum approach too had many elements
similar to the classical approach, with transcribed terminology. He worked
through this over one evening. The next morning when he brought his notes and
explained the formalism to Wolf, he was so pleased that he said he must write
out the paper there and then. So the paper was written out, edited and typed,
and it was only after it had been put in express mail that Sudarshan was
allowed to go for lunch. \ \

Glauber had introduced coherent states to define quantum correlations in
photon detection statistics [\ref{ref:glauberprl}]. However as discussed in
[\ref{ref:mehta}] and [\ref{ref:simonmds}] this proposal was too broad, and
did not have the specificity and advantages of the Diagonal representation to
be discussed below. The timely paper by Sudarshan[\ref{ref:ecgprl}] introduced
the sharply defined Diagonal representation as a generalisation of the
classical correlation functions, and gave the complete answer to the question
of radiation intensity measurements using coherent states. Within the next few
months the next paper of Glauber on this topic appeared, proposing a formalism
on the lines of the Diagonal representation, dubbed
P-representation[\ref{ref:glauberpr131}], however the treatment could be seen
to be incomplete in several technical aspects [\ref{ref:mehta}]. The contrast
in the extent of contribution of the two authors, and the gap in insight and
clarity as to the final synthesis are traced out in [\ref{ref:simonmds}].
Specifically the priority regarding the Diagonal representation, also known as
Optical Equivalence Theorem belongs to George Sudarshan. It provided the sharp
criteria identifying what are non-classical states of radiation that inspired
new experiments [\ref{ref:kimble}]. It is acknowledged in the literature as
the Sudarshan-Glauber, though sometimes also the Glauber-Sudarshan
[\ref{ref:wolfmandel}], representation. \

\section{Recovering uniqueness}

The great redundancy in the basis $\{ | z \rangle \langle z' | \}$ however
has a very simple and elegant resolution as was realised by E. C. G. Sudarshan
[\ref{ref:ecgprl}]. Accordingly, it is sufficient to represent any density
matrix as the diagonal entity
\[ \rho = \int d^2 z \phi (z) | z \rangle \langle z | \nobracket \]
The quantity $\phi (z)$ is then unique unlike $F (z, z')$, however it is not
restricted to be an ordinary function. In general it turns out to be a
distribution, or a ``generalised function'', of which the well known Dirac
$\delta$-function is perhaps the simplest example. In his 1963 paper Sudarshan
gave a conversion formula that expresses $\phi (z) \equiv \phi (r e^{i
\theta})$ in terms of the standard $n$ representation as
\begin{eqnarray*}
  \phi (z) & = & \sum_{n, n'} \frac{\sqrt{n!n' !} }{(n + n') !2 \pi r} \rho_{n
  n'} \times\\
  &  & \exp \{ r^2 + i (n' - n) \theta \} \left( - \frac{\partial}{\partial
  r} \right)^{n + n'} \delta (r)
\end{eqnarray*}
We may well feel great uneasiness at the high order of derivative of the
$\delta$ function involved. Nevertheless, this is the complete explicit answer
and appeared in literature before the P-representations made their appearance.

In the quantum theory one defines the second order coherence function as
\[ \Gamma (x_1, x_2) = \tmop{Tr} \{ E (x_1) \rho E^{\dag} (x_2) \} \]
where $x$ stand for space as well as time coordinates and $E$ and $E^{\dag}$
are the positive and negative frequency parts of the electric field operator.
The usual expression for $E$ is sum over all positive frequency mode
functions. To avoid complication we stay with the single mode system we have
been using, in terms of which this becomes
\[ \Gamma = \tmop{Tr} (a \rho a^{\dag}) \equiv \langle a^{\dag} a \rangle =
   \int d^2 z \phi (z) z^{\ast} z \]
where $\phi$ corresponds to the density matrix $\rho$ defining the averaging
$\langle \rangle$. Thus the coherence function has the appearance of a generic
classical quantity, the average value of a function of the complex variable
$z$ with statistical distribution given by $\phi (z)$. The main difference is
that $\phi$ can be the rather wild object without any guarantee of positivity
or even of validity as a usual function. But it does reexpress the quantum
coherence function uniquely as an expression with direct analogy to the
classical expression. Further it can be shown that these coherence functions
also have the convexity property. Finally corresponding to the unimodular
reduced coherence functions which act as generators in the classical
formalism, here there are states with excitation of the field to the required
degree in a single mode. To quote the originator of the formalism
([\ref{ref:ecgreview}] pg. 138, original emphasis retained),

\begin{quotation}
  ``{\tmem{Thus all the results of modern classical theory of second order
  partial coherence are unaltered in the quantum theory formulation.''}}
\end{quotation}

\section{A resolution through the unresolved }

In hindsight, the disagreement between the Newtonian and Huygens schools on
the nature of light appears rather innocent. The concepts of particles and
waves were rather clear cut then, and the issue was simply of identifying
light as one or the other. By comparison, now we seem to have the final theory
of light in hand, thanks to the developments leading from Planck and Einstein,
to Bose, and finally to Sudarshan. Yet the answer may leave us more perplexed
than when we did not know so much. The reason is that the concepts in terms of
which we achieved this synthesis are themselves rather novel, and the
comprehensive theory of light now lives in a domain which we will not be able
to report successfully to general public. In other words, if a high school
student asks us, is the controversy now settled? The answer is yes. But if the
next question is, can you tell me which way it got settled? The answer is no.
\ \

A century after the development of quantum theory, all of the lay population
and substantial segments of the professional community remain puzzled if not
befuddled by the principles of the new mechanics. If the wave particle duality
was not puzzling enough, the outcome of an act of measurement is statistical
in nature. The mischief however started with Einstein's own most crisply
articulated photon hypothesis quoted at the beginning of the article, \ ``...
a light ray emitted from a point source, ... consists of finite number of
energy quanta localised at points of space ...''. One is tempted to ask, which
points of space? The moment a {\tmem{point}} source chooses one or more
specific directions in space into which to send out the emission, it is
violating rotational invariance. We know the answer with the hindsight of
innumerable experiments. The isotropy of the process is recovered
statistically, after a sufficiently large number of emissions has been
observed. Einstein inadvertently but with deep insight had already introduced
the drastic new element of quantum theory.

As to the wave particle duality, Dirac alone among all the stalwarts seems to
have stood by the new positive principle in quantum mechanics, that of linear
superposition principle. And once one accepts this principle, most puzzles
over ``dual'' description vanish. And the principle plays an important role in
this new synthesis in the theory of light. But this leads us to ponder another
striking fact. If all the results of classical coherence theory are subsumed
in the fully quantum formulation then we may conjecture that the linear
superposition principle observed in many of the electromagnetic phenomena we
daily use and control may well have been inherited from quantum mechanics.

\section*{Acknowledgement}

The author wishes to thank Professor N. Mukunda and Professor Xerxes Tata for
critical comments on the manuscript.

\begin{tmornamented}
 \begin{center}
\includegraphics[width=0.4\columnwidth]{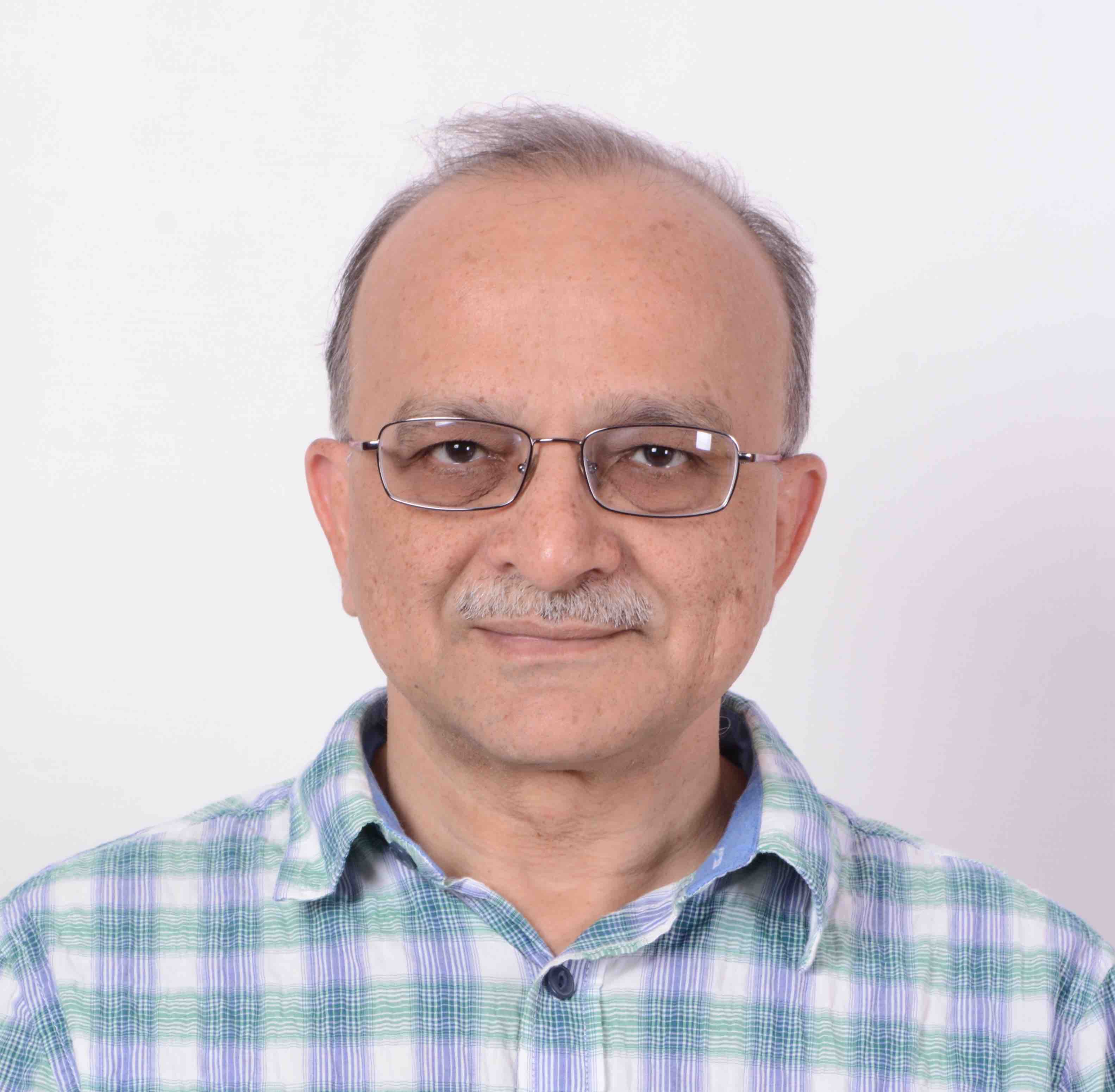}
  \end{center}
  
  Urjit A. Yajnik obtained masters in physics from IIT Bombay and PhD in
  physics from the University of Texas at Austin in 1986, under the
  supervision of Prof. E. C. G. Sudarshan. After a postdoctoral position at
  TIFR, he has been on the faculty of IIT Bombay since 1989. He has been a
  visiting professor at McGill University and the Universite de Montreal
  Canada, and University of California Irvine, USA. He has been a visitor to
  ICTP Trieste, Italy, Perimeter Institute, Canada and other centres of
  advanced research in Germany, S. Korea, Japan and China. His research
  interests are unified theories, supersymmetry, general theory of relativity
  and cosmology.
\end{tmornamented}

\section*{References}

\begin{enumerate}
  \item Urjit A. Yajnik\label{ref:uayphotonsonetwo}, ``The Conception of
  photons'' Parts I and II, {\tmem{Resonance}} (2015) 1085 and (2016) 49.
  
  \item E. C. G. Sudarshan\label{ref:ecgreview}, ``Quantum Theory of Partial
  Coherence'', {\tmem{J. Math. Phys. Sci.}} {\tmstrong{3}}, (1969) 121
  
  \item E. Merzbacher, {\tmem{Quantum Mechanics}}, Wiley Eastern. (Either of
  2nd or 3rd ed.s)\label{ref:merz}
  
  \item N. Mukunda,\label{ref:mukunda} ``Operator properties of generalised
  coherent states'', {\tmem{Pramana}}, {\tmstrong{56}} (2001) 245
  
  \item C. L. Mehta,\label{ref:mehta} ``Sudarshan Diagonal Representation :
  Developments and Applications'', Sudarshan Seven Quests Symposium, {\tmem{J.
  Phys. Conference Series}} {\tmstrong{196}}, (2009) \ 012014
  
  \item ``Symmetry and Mathematics\label{ref:ecgftof} : pioneering insights
  into the structure of Physics; E. C. G. Sudarshan talks to Urjit A.
  Yajnik'', {\tmem{Resonance}} (2015) 0264
  
  \item R. J. Glauber\label{ref:glauberprl}, {\tmem{Phys. Rev. Lett.}}
  {\tmstrong{10}} (1963) 84
  
  \item R. Simon and M. D. Srinivas\label{ref:simonmds}, ``Sudarshan's
  diagonal representation : the ecstacy and agony of another major discovery
  in science'', Sudarshan Seven Quests Symposium, {\tmem{J. Phys. Conference
  Series}} \ {\tmstrong{196}}, (2009) \ 012016
  
  \item E. C. G. Sudarshan\label{ref:ecgprl}, {\tmem{Phys. Rev. Lett.}}
  {\tmstrong{10}} (1963) 277
  
  \item R. J. Glauber\label{ref:glauberpr131}, {\tmem{Phys. Rev.
  }}{\tmstrong{131}} (1963) 2766
  
  \item H. J. Kimble\label{ref:kimble}, ``Neoclassical light -- An assessment
  of the voyage into Hilbert space'' Sudarshan Seven Quests Symposium,
  {\tmem{J. Phys. Conference Series}} \ {\tmstrong{196}}, (2009) \ 012015
  
  \item E. Wolf and L. Mandel\label{ref:wolfmandel} {\tmem{Optical Coherence
  and Quantum Optics}}, Cambridge University Press 1995
\end{enumerate}

\

\

\

\
\end{document}